# Real-time observation of bond-by-bond interface formation during oxidation of H-terminated (111)Si

Bilal Gokce[1*], Eric J. Adles[1,2*], David E. Aspnes[1], & Kenan Gundogdu[1a]

**Atomic-level structure of solids is typically determined by techniques such as X-ray and electron diffraction,[1, 2, 3, 4] which are sensitive to atomic positions. It is hardly necessary to mention the impact that these techniques have had on almost every field of science. However, the bonds between atoms are critical for determining the overall structure. The dynamics of these bonds have been difficult to quantify. Here, we combine second-harmonic generation and the bond-charge model of nonlinear optics[5, 6] to probe, in real time, the dynamics of bond-by-bond chemical changes during the oxidation of H-terminated (111)Si, a surface that has been well characterized by static methods. We thus demonstrate that our approach provides new information about this exhaustively studied system. For example, oxidation is activated by a surprisingly small applied macroscopic strain, and exhibits anisotropic kinetics with one of the three equivalent back-bonds of on-axis samples reacting differently from the other two. Anisotropic oxidation kinetics also leads to observed transient changes in bond directions. By comparing results for surfaces strained in different directions, we find that in-plane control of surface chemistry is possible. The use of nonlinear optics as a bond-specific characterization tool is readily adaptable for studying structural and chemical dynamics in many other condensed-matter systems.**

Real-time data on the kinetics of different classes of bonds during chemical reactions can provide new insight into the reactions themselves, leading to potentially unprecedented control over interface formation. Here, we show that a combination of second-harmonic generation (SHG) and bond-charge modeling reveals previously unrecognized complexities in the oxidation kinetics of H-terminated (111)Si. For example, an externally applied strain activates then manipulates the oxidation kinetics of bonds, which in turn causes transient structural changes observed as deviations in bond directions.

We used p-polarization for both illumination and detection. The contributions of the individual types of bonds are isolated by rotating the sample and recording the SHG anisotropy (SHGA). Nonlinear-optical data are generally described in terms of macroscopic crystal symmetries.[7] However, during chemical reactions these symmetries break and reform, making tensor-based interpretations challenging. We overcome this difficulty with the anisotropic bond-charge model. When a particular asymmetric bond is aligned parallel to the driving field, the acceleration of its bond charge, and hence its radiated SHG signal, is maximized (see supplementary information). This orientation dependence provides the opportunity to follow chemical changes on a bond-type-specific basis in real time. We track not only the evolution of amplitudes, for which some general data are already available,[8, 9, 10] but also that of average bond directions, which provide new atomic-scale information about strain. The result is an improved picture of oxidation kinetics at the (111)Si interface.

As background, the outermost atoms of the (111) surface of a tetrahedrally bonded material such as Si have one "up" orbital and three Si–Si "back" bonds. For Si, $NH_4F$ treatment passivates the "up" orbital with H, forming a Si–H bond. On Pauling's scale, the electronegativity of H is 2.20 while that of Si is 1.90. Hence all 4 outer-layer bonds are ordered and asymmetric, as required for SHG: the "up" bonds asymmetric from the electronegativity difference itself, and the "back" bonds from chemical induction. In a classical model this asymmetry translates into an anharmonicity of the restoring force acting on the associated bond charge, and thus the efficiency of SHG by the charge. During oxidation O first replaces the H caps then inserts itself into the back bonds, where it forms bridges between Si atoms. Since the electronegativity of O is 3.44, oxidation significantly increases bond asymmetry and hence, particularly in the replacement step, the SHG signal.

Figure 1a is a three-dimensional plot of 113 consecutive SHGA scans at 98 s intervals of an unstrained, initially H-terminated on-axis control sample in air. The data exhibit six features. After a 10-min incubation period, the amplitudes of three of these increase dramatically. These correspond to the successive near (~7.8°) alignments of the back bonds with the internal field of the plane wave. As oxidation proceeds these amplitudes reach a maximum, then decrease. Figure 1b and c shows their evolution, and that of the azimuths at which they occur, as determined by fitting to Gaussian functions. The amplitudes evolve identically and the azimuths remain essentially at their original values, as expected for material with no preferred azimuthal orientation.

We now consider a similar sample but with 0.10% strain along <-12-1>, as determined by reflection anisotropy (RA).[11] This direction is within 19.5° of the <-11-1> bonds, and 61.9° from the other two. The corresponding SHG data are shown in Fig. 2a. Comparing Figs. 1a and 2a, this seemingly small strain significantly changes the oxidation kinetics. First, oxidation starts immediately: there is no incubation period. Second, oxidation is anisotropic: the <-11-1> feature rises more rapidly than the other two. This is seen more clearly in Fig. 2b, which shows that the <-11-1> feature reaches 25% of its maximum amplitude by the end of the first scan and continues to rise rapidly thereafter. However, after 20 min the other features catch up. All three reach their maximum amplitudes at about 66 min then decrease, asymptotically approaching the common value as shown by the green points in Figs. 3a–c.

Elimination of the incubation period indicates immediately that strain enhances H replacement. However, the observed asymmetry of the SHG signal implies a corresponding asymmetry in the oxidation of the back bonds. We note first that the Si–O bond length of about 1.6 Å is incommensurate with the Si–Si bond length of 2.35 Å, so anisotropic oxidation must generate additional anisotropic strain. We detect this through changes of the azimuths of the SHGA peaks. These are also given in Figs. 3a-c for the sample of Fig. 2, and can be compared to the control data shown in Fig. 1c. The red dashed line shows the equilibrium azimuths. Significant discrepancies are seen in the first few minutes: the <1-1-1> and <-1-11> bonds both move toward smaller angles, i.e., closer to the <-11-1> bond. Also, the <-11-1> bond is first rapidly displaced toward <-1-11>. However, with increasing time the bonds move back toward their equilibrium positions. These changes indicate an initial accumulation, then a relaxation, of microscopic strain.

We can understand these data using the model of Fig. 3d. This simulation, based on molecular-mechanics (MM2) force-field calculations, shows the result of inserting an O atom in a <-11-1> bond. When insertion occurs, the affected Si atoms are pushed apart, consistent with the data shown in Figs. 3a and 3c. Thus we conclude that the applied strain enhances the probability of the insertion of O in the bond(s) most nearly parallel to the strain, and that insertion further enhances the SHG signal associated with the affected bond(s). In a purely random process the sums of these deviations should equal zero, which occurs for short and long times although not at the intermediate stage.

Given the now-demonstrated importance of strain on oxidation, we performed experiments with strain applied in different directions both to critically assess the above model and to determine the extent to which bond-level oxidation can be controlled. Figure 4 shows the results of 0.08% strain applied in the <1-10> direction, where it is orthogonal to <-11-1> and makes a 35.3° angle with the <1-1-1> and <-1-11>. All peaks start with the same amplitude, but anisotropy quickly appears as the <1-1-1> and <-1-11> oxidize more rapidly than the <-11-1> bonds. Thus the strain direction does provide some control over oxidation rates. We confirm the more rapid oxidation of the <1-1-1> and <-1-11> bonds by comparing the evolution of the peak azimuths, shown in Figs. 4b-d, with the MM2 model calculation shown in Fig. 4e. The calculation shows that this oxidation should push the <1-1-1> and <-1-11> closer to each other by 3° to 7°, consistent with the results shown in Figs. 4b-d.

To relate our observations to conventional oxidation data, we followed the evolution of both interface and $SiO_2$ thicknesses by spectroscopic ellipsometry (SE) (see supplementary information Fig. S7). We found that the transient structural dynamics observed during the first 80 min are associated with the growth of a 2 to 4 Å thick $SiO_2$ layer, which corresponds to the oxidation of the topmost bilayer of the surface. Our SHGA data allow us to conclude that macroscopic strain reduces with increasing oxide thickness, at which point other factors such as diffusion dominate oxidation kinetics.

Air oxidation of H-terminated (111)Si is known to occur in two stages.[12] First, the top H is replaced with OH, and second, the back-bonds oxidize to form Si–O–Si. The former is nominally the rate-limiting step.[12] However, our SHGA data show that both are kinetically limited on similar time scales, and that an externally applied strain affects both in different ways. This leads to some control over the oxidation process and provides additional insight into the oxidation dynamics. With strain applied, oxidation initially occurs more rapidly in the strain direction, but as oxidation progresses eventually all directions become equivalent. The final long-term decrease in SHG intensity is also a result of further oxidation, because in the $SiO_2$ limit the SHG contribution of the oxidized material vanishes by directional disorder, and is replaced by that of the next bilayer down.

We have also performed measurements on vicinal surfaces to verify that steps are not a factor. The initial-oxidation kinetics of vicinal surfaces are strikingly different from those of either strained and unstrained on-axis surfaces (see supplementary information), and will be reported elsewhere.

While structural changes in condensed-matter systems have long been probed by methods such as X-ray and electron diffraction, we show that nonlinear optics can be used to probe analogous changes in bond structure. In the specific case studied here, we obtain information about the Si oxidation in unprecedented detail. By comparing measurements on strained and unstrained on-axis samples, we find that strain has a surprisingly large influence on kinetics, and hence in principle can be used to develop methods that enable unprecedented in-plane control of surface chemistry, for example to produce interfaces engineered at the bond level. We are continuing to investigate the physics of the strain-induced anisotropic oxidation, which will be an important step in this direction. Our approach is general, and can be applied to problems ranging from bond formation during chemical changes on surfaces to bond dynamics in functional materials that exhibit phase transitions.

# Figure Captions:

Figure 1. a) Evolution of SHGA p-p data for an unstrained, initially H-terminated on-axis (111)Si control sample during air exposure. The dominant features correspond to the approximate alignment of the <-11-1>, <-1-11>, and <1-1-1> bonds with the internal driving field. b and c) Evolution of the amplitudes and azimuths of the dominant features.

Figure 2: a) As Fig. 1, but with 0.10% strain along <1-21>. b) The first four SHGA scans. The <-11-1> feature rises more rapidly than the other two.

Figure 3: a,b,c) Blue points: evolution of the azimuths of the three major features. Dashed red lines: azimuths for the unperturbed crystal. Green points: corresponding amplitudes. d) Atomic-scale diagram indicating expected bond distortions with O inserted in the <-11-1> bond.

Figure 4: a) As Figs. 1a and 2a but with 0.08% strain along <1-10>. b,c,d) As Fig. 3a,b,c. e) As Fig. 3d but with O inserted in the <-1-11> and <1-1-1> bonds.

# Methods:

Data-acquisition and analysis methods are discussed in the supplementary information.

Sample Preparation: Two different sets of (111)Si samples were used, one oriented at 0.0 ± 0.1° and the other at 4.6 ± 0.1° toward [11-2] as determined by X-ray diffraction (XRD). Each sample investigated was divided into two pieces for SHG and SE measurements. Samples were cleaned by consecutive 10-min immersions in 80 °C $NaOH/H_2O_2/H_2O$ (1:1:5) and 80 °C $HCl/H_2O_2/H_2O$ (1:1:5). Native oxides were then stripped and the samples capped with H by a 20-min immersion in 40% $NH_4F$. To prevent pitting the $NH_4F$ solution was deoxygenated prior to immersion.[13] Measurements were begun approximately 90 s after $NH_4F$ emersion, after the surfaces were dried with high-purity $N_2$.

Figure 1

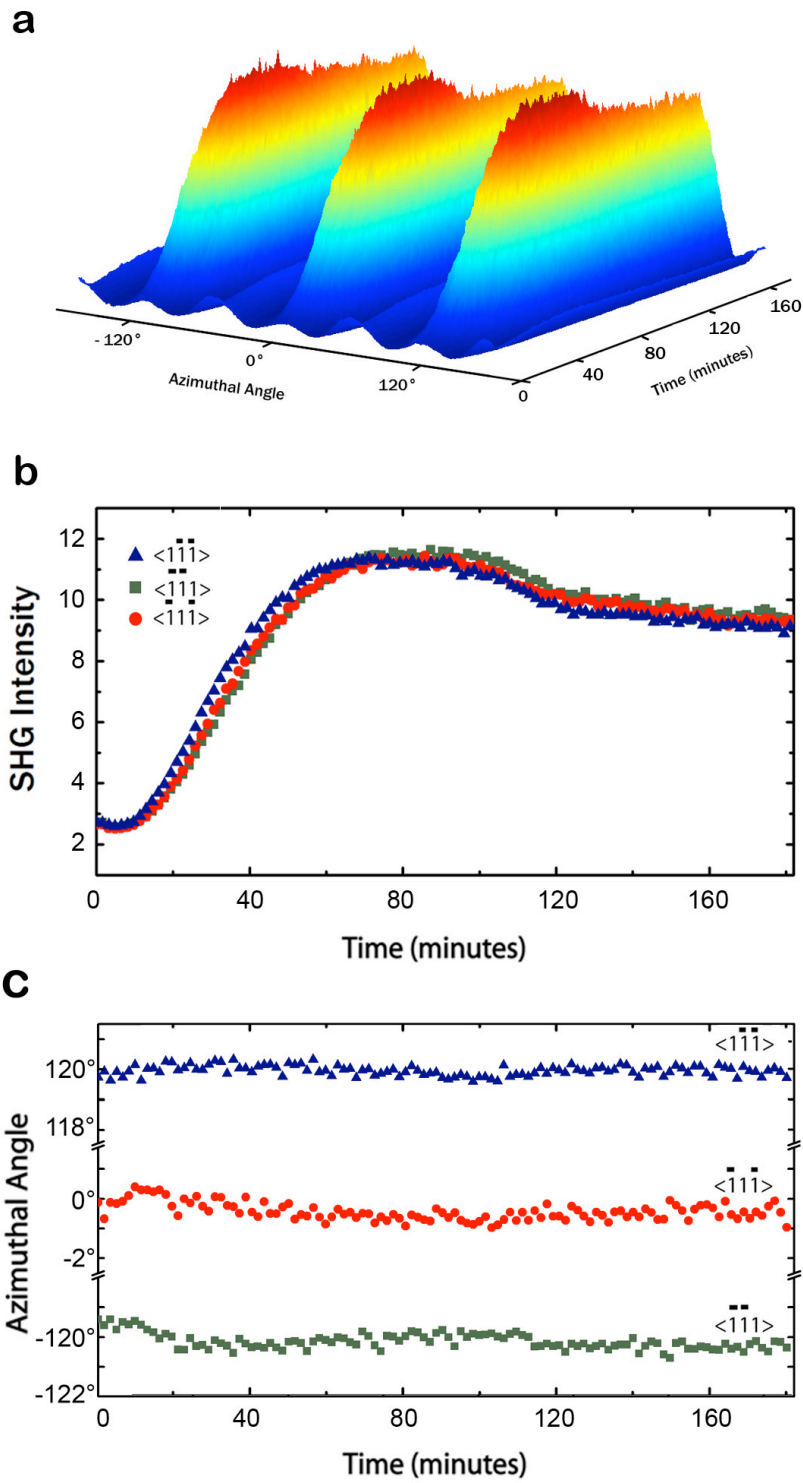

Figure 2

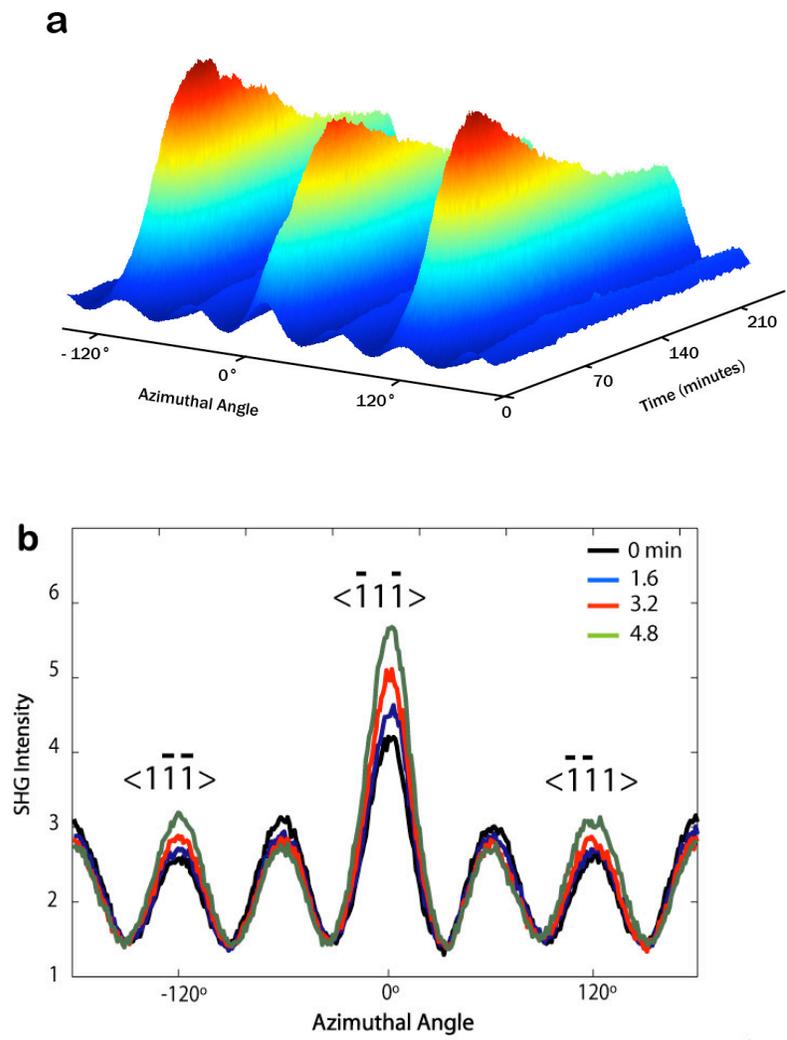

Figure 3

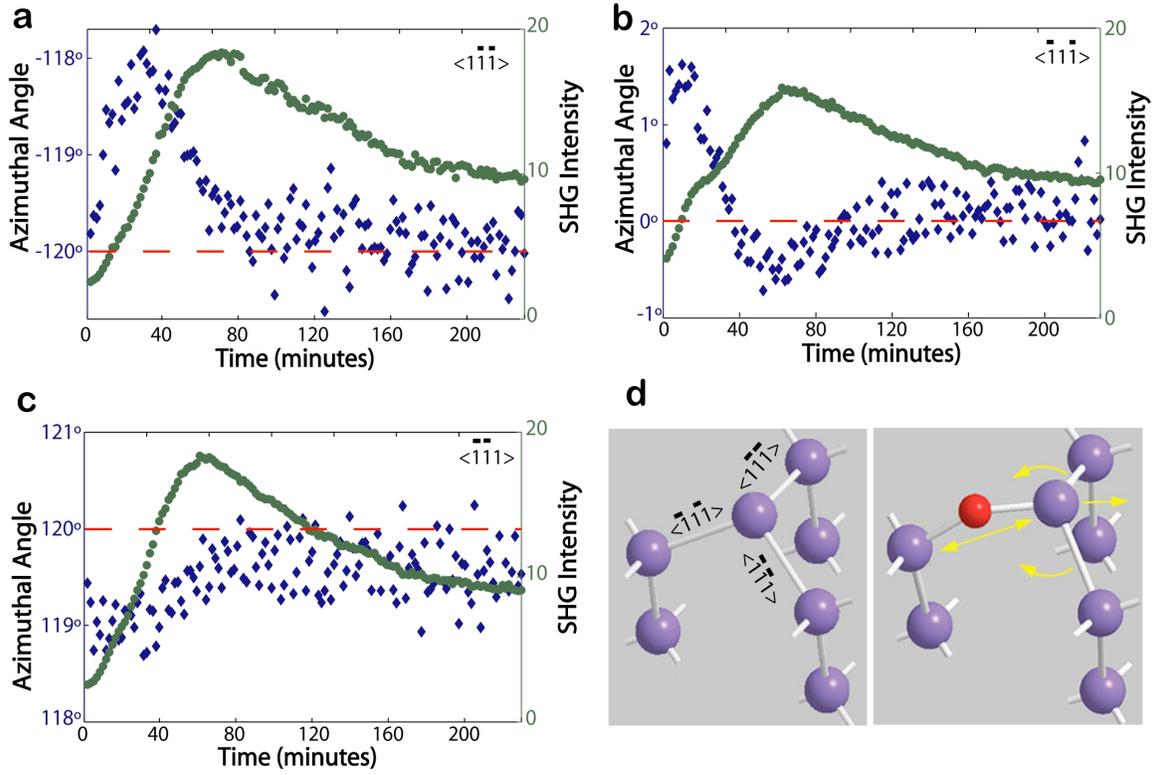

Figure 4

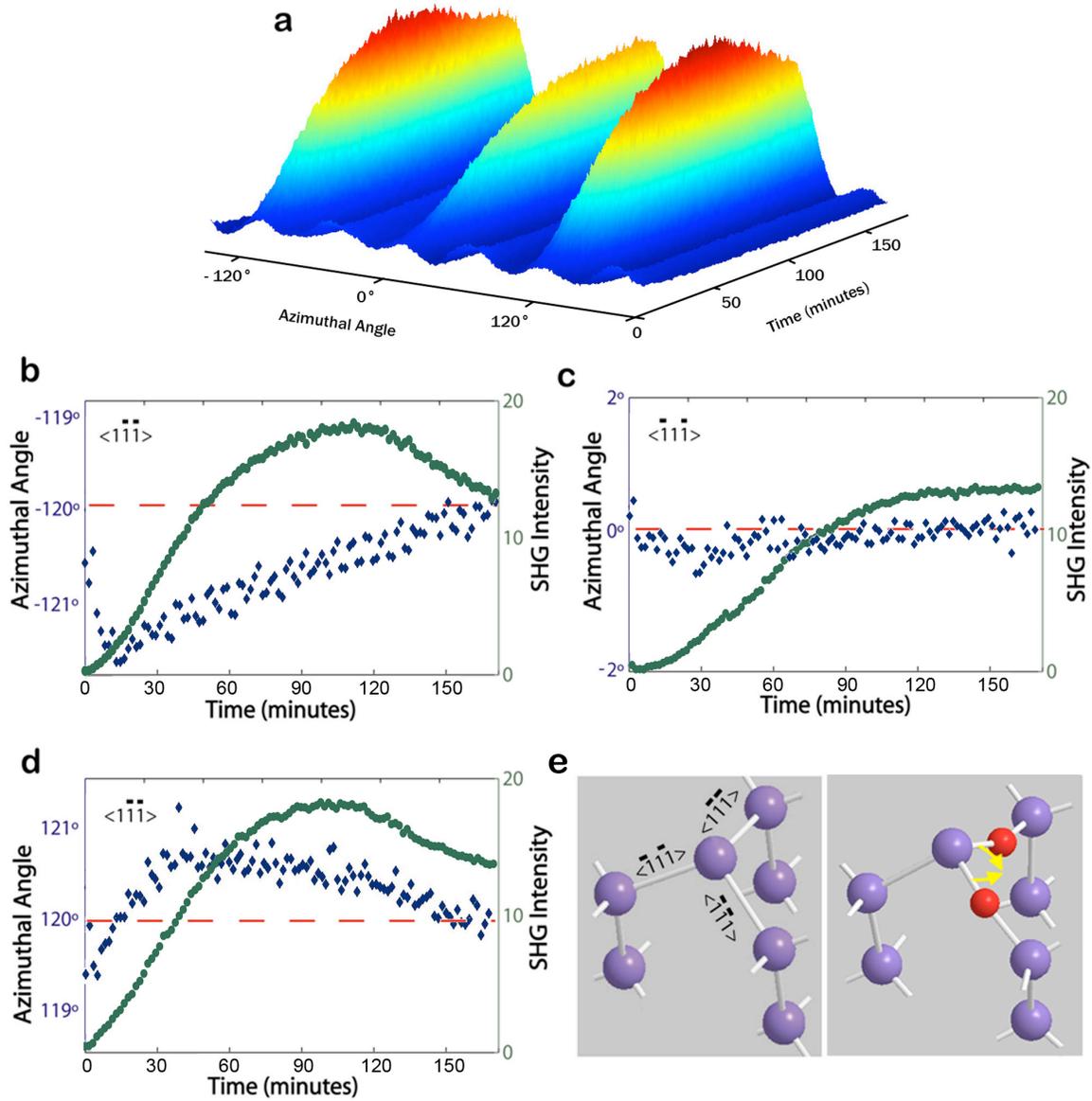

# Supplementary Information

Supplementary information accompanies this paper.

# Competing Interests Statement

The authors declare no competing financial interests

# Author Information

[1] Physics Department, North Carolina State University, Raleigh, NC 27695-8202

[2] Center for Advanced Studies in Photonics Research, University of Maryland Baltimore County, Baltimore, MD, 21250

[*] These authors contributed equally to this work.

[a] Correspondence to: K. Gundogdu[1] Correspondence and requests for materials should be addressed to K.G (Email: kgundog@ncsu.edu)